\documentclass[11pt,fleqn]{article}

\usepackage{amsmath,amsfonts,amssymb}
\usepackage{latexsym}
\usepackage{amsmath} 
\usepackage{amssymb} 
\usepackage{bm}
\usepackage{graphicx}
\usepackage{subfigure}

\begin{document} 


\title{Mach like principle from conserved charges}
\date{\today}
\author{Eduardo Guendelman \and Roee Steiner \\ Physics Department, Ben-Gurion University of the Negev, Beer-Sheva 84105, Israel}

\maketitle

\abstract{
We study models where the gauge coupling constants,  masses and the gravitational constant are functions of some conserved charge in the universe, and furthermore a cosmological constant that depends on the total charge of the universe. We first consider the standard Dirac action, but where the mass and the electromagnetic coupling constant are a function of the charge in the universe and afterwards extend this to curved spacetime and consider gauge coupling constants, the gravitation constant and the mass as a function of the charge of the universe, which represent a sort of Mach principle for all the constants of nature.
In the flat space formulation, the formalism is not manifestly Lorentz invariant, however Lorentz invariance can be restored by performing a phase transformation of the Dirac field, while in the curved space time formulation, there is the additional feature that some of the equations of motion break the general coordinate invariance also, but in a way that can be understood as a coordinate choice only, so the equations are still of the General Relativity type, but with a certain natural coordinate choice, where there is no current of the charge.
We have generalized what we have done and also constructed a cosmological constant which depends on the total charge of the universe.
If we were to use some only approximately conserved charge for these constructions, like say baryon number (in the context of the standard model), this will lead to corresponding violations
of Lorentz symmetry in the early universe for example.
We also briefly discuss another no local formulations where the
coupling constants are  functions of the Pontryagin index of some non abelian gauge field configurations.
The construction of charge dependent contributions can also be motivated from the structure of the "infra-red counter terms" needed to cancel infra red divergences for example in three dimension.
}

\section{Introduction}
Mach's principle is well known, as a principle that relates a local problem to a non local problem. The original Mach principle \cite{Mach} is based on the claim that the inertial frames are influenced by the other celestial bodies.
In other words' every mass in the universe is influenced by all the other masses in the universe.
The Mach principle is still in a debate. 
In our article, we will show that there is a possibility to precisely formulate a Mach principle  for electromagnetic coupling constant and indeed any other constant of nature, where we take these constants to be a function of the total charge. This breaks the locality of the problem (in the original Mach principle the mass broke the locality). We briefly discuss also the possibility of these constants depending on the Pontryagin index of some non abelian gauge field configurations.\\
In ref.\cite{jackiw2} ,in order to cancel infra red divergence in 3  - dimension gauge theories , infra red counter terms are introduced. This is shown in ref.\cite{jackiw2} to be equivalent to the procedure developed in ref\cite{guend} to cancel infra red divergence by introducing zero energy momentum photons.
These infra red divergences are related to the super renormalizabilty of the theory in 3- dimensions \cite{jackiw3}
We generalize this idea and define the coupling constant to be proportional to this term (and latter to be an arbitrary function of (\ref{eq: counter term})), we latter connect this procedure with our treatment.
 
\section{The electromagnetic coupling constant as a function of all the charge in the universe}\label{per:charge}
We begin by considering the action for the Dirac equation (see for example ref \cite{basic})
\begin{equation}\label{eq:simple dirac action}
S=\int d^{4}x\, \bar{\psi}(\frac{i}{2}\gamma^{\mu}\stackrel{\leftrightarrow}{\partial}_{\mu}-eA_{\mu}\gamma^{\mu}-m)\psi
\end{equation}
  where $ \bar{\psi}=\psi^{\dagger}\gamma^0 $.However here we take the coupling constant $ e $ to be proportional to the total charge (we will afterwards generalize and consider an arbitrary function of the total charge).

\begin{equation}\label{eq:def e}
e=\lambda\int\psi^{\dagger}(\vec{y},y^{0}=t_{0})\psi(\vec{y},y^{0}=t_{0})\, d^{3}y=\lambda\int \rho(\vec{y},y^{0}=t_{0})\, d^{3}y
\end{equation}

and we will show that physics does not depend on the time slice $ y^{0}=t_{0} $

so after the new definition of "e" the  action will be:

\begin{eqnarray}
S=\int d^{4}x\, \bar{\psi}(x)(\frac{i}{2}\gamma^{\mu}\stackrel{\leftrightarrow}{\partial}_{\mu}-m)\psi(x)\nonumber\\-\lambda(\int d^{3}y\, \bar{\psi}(\vec{y},y^{0}=t_{0})\gamma^{0}\psi(\vec{y},y^{0}=t_{0}))(\int d^{4}x\, \bar{\psi}(x)A_{\mu}\gamma^{\mu}\psi(x))
\end{eqnarray}

we can express the three dimensional integral as a four dimensional integral

\begin{equation}\label{eq:trick}
\int d^{3}y\, \bar{\psi}(\vec{y},y^{0}=t_{0})\gamma^{0}\psi(\vec{y},y^{0}=t_{0})= \int d^{4}y\, \bar{\psi}(y)\gamma^{0}\psi(y)\delta(y^{0}-t_{0}) 
\end{equation}

so finally the action will be

\begin{equation}
S=\int d^{4}x\, \bar{\psi}(x)(\frac{i}{2}\gamma^{\mu}\stackrel{\leftrightarrow}{\partial}_{\mu}-m)\psi(x)-\lambda(\int d^{4}x\, \bar{\psi}(x)A_{\mu}\gamma^{\mu}\psi(x))(\int d^{4}y\, \bar{\psi}(y)\gamma^{0}\psi(y)\delta(y^{0}-t_{0}))
\end{equation}

if we consider the fact that $ \frac{\delta\bar{\psi}_{a}(x)}{\delta\bar{\psi}_{b}(z)}=\delta^{4}(x-z)\,\delta_{ab} $ and $ \frac{\delta\psi(x)}{\delta\bar{\psi}(z)}=0 $ we get the equation of motion

\begin{eqnarray}
\frac {\delta S}{\delta\bar{\psi}(z)}=0=\int{\delta^{4}(x-z)(i\gamma^{\mu}\partial_{\mu}-m)\psi (x)}\,d^{4}x\,\nonumber\\-\lambda(\int d^{4}x\, \delta^{4}(x-z)A_{\mu}\gamma^{\mu}\psi(x))(\int d^{4}y\, \bar{\psi}(y)\gamma^{0}\psi(y)\delta(y^{0}-t_{0}))\nonumber\\-\lambda(\int d^{4}x\, \bar{\psi}(x)A_{\mu}\gamma^{\mu}\psi(x))(\int d^{4}y\, \delta^{4}(y-z)\gamma^{0}\psi(y)\delta(y^{0}-t_{0}))
\end{eqnarray}

so to accomplish our goal we need just to perform the integrations in the last equation, and then the expression will simplified to

\begin{eqnarray}
\frac {\delta S}{\delta\bar{\psi}(z)}=(i\gamma^{\mu}\partial_{\mu}-m)\psi(z)-\lambda(\int{\bar{\psi}(y)\gamma^{0}\psi(y)\delta(y^{0}-t_{0})}\,d^{4}y)A_{\mu}\gamma^{\mu}\psi(z)&\nonumber\\-\lambda(\int{\bar{\psi}(x)A_{\mu}\gamma^{\mu}\psi(x)}\,d^{4}x)\gamma^{0}\psi(z)\delta(z^{0}-t_{0})
\end{eqnarray}

which can be simplified more by the use of new definition  $ b_{e}=\lambda(\int{\bar{\psi}(x)A_{\mu}\gamma^{\mu}\psi(x)}\,d^{4}x) $ which is a constant, and by the definition in equation (\ref{eq:def e}) 

\begin{equation}\label{eq:motion e}
\frac {\delta S}{\delta\bar{\psi}(z)}=[i\gamma^{\mu}\partial_{\mu}-m-eA_{\mu}\gamma^{\mu}-b_{e}\gamma^{0}\delta(z^{0}-t_{0})]\psi(z)=0
\end{equation}

so we can see that the last term in the equation of motion (\ref{eq:motion e}) contains $ A^{GF}_{\mu}\gamma^{\mu} $ where $ A^{GF}_{\mu}=\partial_{\mu}\Lambda $ and $ \Lambda=b_{e}\theta(z^{0}-t_{0}) $ is a pure gauge field. so the solution of this equation is
\begin{equation}
\psi=e^{-ib_{e}\theta(z^{0}-t_{0})}\psi_{D} 
\end{equation}

  where $ \psi_{D} $ is the solution of the equation
  \begin{equation}
 [i\gamma^{\mu}\partial_{\mu}-m-eA_{\mu}\gamma^{\mu}]\psi_{D}=0 
  \end{equation}
  
    from which it follows that $ j^{\mu}=\bar{\psi}_{D}\gamma^{\mu}\psi_{D}=\bar{\psi}\gamma^{\mu}\psi $ satisfies the local conservation law $ \partial_{\mu}j^{\mu}=0 $ and therefore we obtain that $ Q=\int{d^{3}x\, j^{0}} $ is conserved, so it does not depend on the time slice, furthermore it also follows that it is a scalar as we have proved in the appendix (\ref{appendix Q}).

\section{Mass as a function of all the charge in the universe}
We will show now that we can do the same thing as in paragraph (\ref{per:charge}) for the mass.
We consider the action equation (\ref{eq:simple dirac action}) where we take the mass to be equal to

\begin{equation}\label{eq:def m}
m=\lambda\int{\bar{\psi}\gamma^{0}\psi \,d^{3}y}
\end{equation}

we do the same thing like in paragraph (\ref{per:charge}) so we expand equation (\ref{eq:def m}) like we did in equation (\ref{eq:trick}), so we will get the equation of motion  

\begin{equation}\label{eq:motion m}
\frac {\delta S}{\delta\bar{\psi}(z)}=[i\gamma^{\mu}\partial_{\mu}-m-eA_{\mu}\gamma^{\mu}-b_{m}\gamma^{0}\delta(z^{0}-t_{0})]\psi(z)=0
\end{equation}

where $ b_{m}=\lambda\int{\bar{\psi}\psi \,d^{4}x} $. So again we can eliminate $ \delta(z^{0}-t_{0}) $ term by a phase transformation, with the same conclusion that there is no violation of Lorentz invariance.

\section{Coupling constant as a general function of all the charge in the universe}\label{se: Coupling constant as a general function of all the charge in the universe}

For a general coupling constant $"a"$ and a general function $ F(\bar{\psi},\psi) $ and a coupling constant "e" the action of the Dirac equation is

\begin{equation}\label{eq:general function action}
S=\int d^{4}x\, \bar{\psi}(\frac{i}{2}\gamma^{\mu}\stackrel{\leftrightarrow}{\partial}_{\mu}-eA_{\mu}\gamma^{\mu}-m)\psi+\int{d^{4}x[a F(\bar{\psi},\psi)]}
\end{equation}

we take the coupling constants "a" and "e" as an arbitrary functions $g_{a}$ and $g_{e}$ of

 $Q=\int\psi^{\dagger}(\vec{y},y^{0}=t_{0})\psi(\vec{y},y^{0}=t_{0}) \,d^{3}y $ 

\begin{equation}\label{eq:general}
a=g_{a}(\int\psi^{\dagger}(\vec{y},y^{0}=t_{0})\psi(\vec{y},y^{0}=t_{0}) \,d^{3}y)=g_{a}(\int \rho(\vec{y},y^{0}=t_{0}) \,d^{3}y)
\end{equation}

\begin{equation}
e=g_{e}(\int\psi^{\dagger}(\vec{y},y^{0}=t_{0})\psi(\vec{y},y^{0}=t_{0}) \,d^{3}y)=g_{e}(\int \rho(\vec{y},y^{0}=t_{0}) \,d^{3}y)
\end{equation}

we put the last definition in equation (\ref{eq:general function action}) and expand equation (\ref{eq:general}) like we did in equation (\ref{eq:trick}), we do a variation and use the fact that $ \frac{\delta\bar{\psi}_{a}(x)}{\delta\bar{\psi}_{b}(z)}=\delta^{4}(x-z)\,\delta_{ab} $ and $ \frac{\delta\psi(x)}{\delta\bar{\psi}(z)}=0 $ and do the integration as like we did in peragraph (\ref{per:charge}) so we get to the general motion equation
 
\begin{equation}\label{eq:motion general}
\frac {\delta S}{\delta\bar{\psi}(z)}=[i\gamma^{\mu}\partial_{\mu}-m-eA_{\mu}\gamma^{\mu}-b_{ae}\gamma^{0}\delta(z^{0}-t_{0})]\psi(z)+a(\frac{\partial F(\bar{\psi(z)},\psi(z))}{\partial\bar{\psi}(z)})=0
\end{equation}

where $ b_{ae}=\frac{\partial g_{a}(Q)}{\partial Q} \int{d^{4}x\,F(\bar{\psi},\psi)}+\frac{\partial g_{e}(Q)}{\partial Q}\int{d^{4}x\,\bar{\psi}\gamma_{\mu}\psi A^{\mu}} $
so we get that any coupling constant in this form can be a function of the charge in the universe without violating Lorentz invariance (again after performing the appropriate phase transformation).

\section{Gravitational coupling constant as a function of all the charge in the universe}
We will show that we can define the gravitation coupling constant as a function of all the charge in the universe. We start with the definition of the equations of each of ours parameters because we are dealing with curved space.
The Dirac action in curved space is \cite{dirac rel}

\begin{equation}
S_{D}=\int{d^{4}x \sqrt{g}\bar{\psi}(\frac{i}{2}\gamma^{\alpha}\stackrel{\leftrightarrow}{D}_{\alpha}-m)\psi}
\end{equation}  

where $ D_{\alpha}=e^{\,\,\mu}_{\alpha}\partial_{\mu}-\frac{i}{4}e^{\,\,\mu}_{\alpha}\eta_{\alpha c} \omega^{c}_{ b\mu}\sigma^{\alpha b} $ is the covariant derivative for fermion and $ e^{\mu}_{\, \alpha} $ is the vierbein and $ \sigma^{\alpha \beta}=\frac{i}{2}[\gamma^{\alpha},\gamma^{\beta}] $ is the commutator of the Dirac gamma metric, and the spin connection is $ \omega^{c}_{b\mu}=e^{c}_{\nu}\partial_{\mu}e^{\nu}_{b}+e^{c}_{\nu}e^{\sigma}_{b}\Gamma^{\nu}_{\sigma\mu} $ where $ \Gamma^{\nu}_{\sigma\mu} $ is the Christoffel symbol and $ \eta_{\alpha c} $ is the lorentzian metric.
the second definition is the action of the gravity \cite{grav}

\begin{equation} \label{eq: coordinate action}
S_{G}=-\frac{1}{16\pi G}\int{\sqrt{g}R(x)\,d^{4}x}
\end{equation}

where we will take gravitational constant to be

\begin{equation}\label{eq:def G}
-\frac{1}{16\pi G}=\lambda\,\int{\sqrt{g}\bar{\psi}e^{\,\,0}_{\alpha} \gamma^{\alpha}\psi \,d^{3}y}
\end{equation}

 the total action is

\begin{equation}\label{eq:total G action}
S_{t}=S_{G}+S_{D}
\end{equation}

we will do now the variation on (\ref{eq:total G action}) by the vierbein and by $ \bar{\psi} $:

\begin{eqnarray} \label{eq: veration of gravity}
\delta S_{t}=\frac{1}{8\pi G}\int{\sqrt{g}[R^{\lambda}_{\,\,\mu}-\frac{1}{2}\delta^{\lambda}_{\,\,\mu}R]e^{\alpha}_{\,\,\lambda}\delta e_{\alpha}^{\,\,\mu}\,d^{4}x}+\nonumber\\\lambda(\int{\sqrt{g}R(x)\,d^{4}x}){\int{\bar{\psi}\gamma^{\beta}\psi[e_{\beta}^{\,\,0}e^{\alpha}_{\,\,\mu}\delta e^{\,\,\mu}_{\alpha}+\delta e^{\,\,0}_{\beta}]\sqrt{g}\,d^{3}y}}+\int{d^{4}x \sqrt{g}\,e^{\alpha}_{\,\,\mu}\delta e^{\,\,\mu}_{\alpha}\bar{\psi}(\frac{i}{2}\gamma^{\beta}\stackrel{\leftrightarrow}{D}_{\beta}-m)\psi}\nonumber\\+\int{d^{4}x \sqrt{g}\,\delta e^{\,\,\mu}_{\beta}\bar{\psi}(\frac{i}{2}\gamma^{\beta}\stackrel{\leftrightarrow}{D}_{\mu})\psi}\nonumber\\+\lambda(\int{\sqrt{g}\delta \bar{\psi} e^{\,\,0}_{\alpha}\gamma^{\alpha}\psi\,d^{3}y})(\int{\sqrt{g}R(x)\,d^{4}x})+\int{d^{4}x\sqrt{g}}\delta\bar{\psi}(i\gamma^{\alpha}D_{\alpha}-m)\psi=0
\end{eqnarray}

because we have two different variations, we can produce two equation. The first equation (from the variation with respect to $ \psi $) is:

\begin{equation} 
\lambda(\int{\sqrt{g}\delta\bar{\psi} e^{\,\,0}_{\alpha} \gamma^{\alpha}\psi\,d^{3}y})(\int{\sqrt{g}R(x)\,d^{4}x})+\int{d^{4}x\sqrt{g}}\delta\bar{\psi}(i\gamma^{\alpha}D_{\alpha}-m)\psi=0
\end{equation}

we will use equation (\ref{eq:trick}) and the fact that $ \frac{\delta\bar{\psi}_{a}(x)}{\delta\bar{\psi}_{b}(z)}=\delta^{4}(x-z)\, \delta_{ab} $ and $ \frac{\delta\psi(x)}{\delta\bar{\psi}(z)}=0 $ and after integration we get:

\begin{equation}\label{eq:motion gravity Dirac}
(i\gamma^{\alpha}D_{\alpha}-m+\lambda e^{\,\,0}_{\alpha}\gamma^{\alpha}\delta(z^{0}-t_{0})(\int{d^{4}x\,R\sqrt{g}}))\psi=0
\end{equation}

which mean that again we have a phase transformation of $ \psi $ that eliminates the $ \delta(z^{0}-t_{0}) $ term.
Before we proceed to the second equation we will notice that the expression $ \delta e^{\,\,\mu}_{\beta}\bar{\psi}(\frac{i}{2}\gamma^{\beta}\stackrel{\leftrightarrow}{D}_{\mu})\psi $ in equation (\ref{eq: veration of gravity}) can be modified to:
\begin{equation} \label{eq:modification}
\delta e^{\,\,\mu}_{\beta}\bar{\psi}(\frac{i}{2}\gamma^{\beta}\stackrel{\leftrightarrow}{D}_{\mu})\psi=\frac{i}{4}\delta e^{\,\,\mu}_{\beta}\bar{\psi}(\gamma^{\beta}\stackrel{\leftrightarrow}{D}_{\mu}+\gamma_{\mu}\stackrel{\leftrightarrow}{D^{\beta}})\psi
\end{equation}

So the second equation is(from the variation with respect to the vierbein):

\begin{eqnarray} \label{eq:ver tetrad Action}
-[R^{\lambda}_{\,\,\mu}-\frac{1}{2}\delta^{\lambda}_{\,\,\mu}R]e^{\alpha}_{\,\,\lambda}=8\pi G\lambda(\int{\sqrt{g}R(x)\,d^{4}x}) \bar{\psi}\gamma^{\beta}\psi\delta(y^{0}-t_{0})[e^{\,\,0}_{\beta} e^{\alpha}_{\,\,\mu}+\delta_{\beta}^{\alpha}\delta_{\mu}^{0}]\nonumber\\+8\pi G e^{\alpha}_{\,\,\mu} \bar{\psi}(\frac{i}{2} \gamma^{\alpha}\stackrel{\leftrightarrow}{D}_{\alpha}-m)\psi + 8\pi G \frac{i}{4}\bar{\psi}(\gamma^{\alpha}\stackrel{\leftrightarrow}{D}_{\mu}+\gamma_{\mu}\stackrel{\leftrightarrow}{D^{\alpha}})\psi
\end{eqnarray}

By contracting equation (\ref{eq:ver tetrad Action}) with $ e_{\alpha\gamma}=e^{\delta}_{\,\,\gamma}\eta_{\alpha\delta} $ and using the fact that $ e^{\alpha}_{\,\,\lambda}e^{\delta}_{\,\,\gamma}\eta_{\alpha\delta}=g_{\lambda\gamma} $ and the fact that $ \bar{\psi}(\frac{i}{2} \gamma^{\alpha}\stackrel{\leftrightarrow}{D}_{\alpha}-m)\psi=-\lambda(\int{\sqrt{g}R(x)\,d^{4}x}) \bar{\psi}e^{\,\,0}_{\beta}\gamma^{\beta}\psi\delta(y^{0}-t_{0}) $ we get the Einstein field equation with a modification 

\begin{eqnarray}\label{eq:Einstein field equation}
R_{\gamma\mu}-\frac{1}{2}g_{\mu\gamma}R=-8\pi G\bar{\psi}(\frac{i}{4} \gamma_{\gamma}\stackrel{\leftrightarrow}{D}_{\mu}+\frac{i}{4} \gamma_{\mu}\stackrel{\leftrightarrow}{D}_{\gamma})\psi\nonumber\\-8\pi \lambda G(\int{\sqrt{g}R(x)\,d^{4}x}) \bar{\psi}\gamma^{\beta}\psi\delta(y^{0}-t_{0})[e^{\,\,\delta}_{\beta} g_{\delta\gamma}\delta^{0}_{\mu}]
\end{eqnarray}

Because equation (\ref{eq:Einstein field equation}) is symmetric (the last term is not symmetric) then we have to conclude that 

\begin{equation}
\bar{\psi}\gamma^{\beta}e^{\,\,\delta}_{\beta}g_{\delta i}\psi=0
\end{equation}

which means that 
\begin{equation}\label{eq: Strike on coordinate}
\bar{\psi}\gamma^{(*)}_{i}\psi=0
\end{equation}
where $ \gamma^{(*)}_{i}=\gamma^{\beta}e^{\,\,\delta}_{\beta}g_{\delta i} $ , which means that we have imposed a gauge condition for the coordinate so that spatial part of the Dirac current is equal to zero, so equation (\ref{eq:Einstein field equation}) will become in these  coordinates,

\begin{eqnarray}\label{eq:Einstein field equation 1}
R_{\gamma\mu}-\frac{1}{2}g_{\mu\gamma}R=-8\pi G\bar{\psi}(\frac{i}{4} \gamma_{\gamma}\stackrel{\leftrightarrow}{D}_{\mu}+\frac{i}{4} \gamma_{\mu}\stackrel{\leftrightarrow}{D}_{\gamma})\psi\nonumber\\-8\pi \lambda G(\int{\sqrt{g}R(x)\,d^{4}x}) \bar{\psi}\gamma^{\beta}e^{\,\,\delta}_{\beta}g_{\delta 0}\psi\delta(y^{0}-t_{0})[\delta^{0}_{\gamma}\delta^{0}_{\mu}]
\end{eqnarray}

It is important to note that $ \gamma $ and $ \mu $ in $ \gamma_{\gamma} $ and $ \gamma_{\mu} $ in equation (\ref{eq:Einstein field equation 1}) are general coordinate indices while $ \beta $ in $ \gamma^{\beta} $ is a Lorentz index.

The solution of equation (\ref{eq:motion gravity Dirac}) is $ \psi=e^{i\lambda(\int{\sqrt{g}R(x)\,d^{4}x})\theta (z^{0}-t_{0})}\psi_{D} $ where $ (i\gamma^{\alpha}D_{\alpha}-m)\psi_{D}=0 $.
If we expand the term $ \bar{\psi}(\frac{i}{4} \gamma_{\gamma}\stackrel{\leftrightarrow}{D}_{\mu}+\frac{i}{4} \gamma_{\mu}\stackrel{\leftrightarrow}{D}_{\gamma})\psi $ in equation (\ref{eq:Einstein field equation 1}) we will get:

\begin{eqnarray}
\bar{\psi}(\frac{i}{4} \gamma_{\gamma}\stackrel{\leftrightarrow}{D}_{\mu}+\frac{i}{4} \gamma_{\mu}\stackrel{\leftrightarrow}{D}_{\gamma})\psi=\bar{\psi}_{D}(\frac{i}{4} \gamma_{\gamma}\stackrel{\leftrightarrow}{D}_{\mu}+\frac{i}{4} \gamma_{\mu}\stackrel{\leftrightarrow}{D}_{\gamma})\psi_{D}\nonumber\\-\frac{\lambda}{2}(\int{\sqrt{g}R(x)\,d^{4}x})\delta(z^{0}-t_{0})\bar{\psi}_{D}[\gamma_{\gamma}\delta^{0}_{\mu}+\gamma_{\mu}\delta^{0}_{\gamma}]\psi_{D}
\end{eqnarray}

If we use the condition in equation (\ref{eq: Strike on coordinate}) we will get:
\begin{eqnarray}
\bar{\psi}(\frac{i}{4} \gamma_{\gamma}\stackrel{\leftrightarrow}{D}_{\mu}+\frac{i}{4} \gamma_{\mu}\stackrel{\leftrightarrow}{D}_{\gamma})\psi=\bar{\psi}_{D}(\frac{i}{4} \gamma_{\gamma}\stackrel{\leftrightarrow}{D}_{\mu}+\frac{i}{4} \gamma_{\mu}\stackrel{\leftrightarrow}{D}_{\gamma})\psi_{D}\nonumber\\-\lambda(\int{\sqrt{g}R(x)\,d^{4}x})\delta(z^{0}-t_{0})\bar{\psi}_{D}\gamma_{0}[\delta^{0}_{\gamma}\delta^{0}_{\mu}]\psi_{D}
\end{eqnarray}

so equation (\ref{eq:Einstein field equation 1}) will become to the ordinary Einstein equation with the conventional form for the energy momentum tensor for fermions:

\begin{eqnarray}\label{eq:Einstein field equation 3}
R_{\gamma\mu}-\frac{1}{2}g_{\mu\gamma}R=-8\pi G\bar{\psi}_{D}(\frac{i}{4} \gamma_{\gamma}\stackrel{\leftrightarrow}{D}_{\mu}+\frac{i}{4} \gamma_{\mu}\stackrel{\leftrightarrow}{D}_{\gamma})\psi_{D}
\end{eqnarray}

since both sides of equation (\ref{eq:Einstein field equation 3}) have a nice covariant structure the same form of equation (\ref{eq:Einstein field equation 3}) will be maintained in any coordinate system.

\section{Gravitational coupling constant as a general function of all the charge in the universe}\label{Gravitational coupling constant as a general function of all the charge in the universe}
We will redefine equation (\ref{eq:def G}) to be:

\begin{equation}\label{eq:def G 2}
-\frac{1}{16\pi G}=\lambda\,F(\int{\sqrt{g}\bar{\psi}e^{\,\,0}_{\beta} \gamma^{\beta}\psi \,d^{3}y})=\lambda\,F(Q)
\end{equation}

so now equation (\ref{eq: veration of gravity}) will be:

\begin{eqnarray} \label{eq: veration of gravity 2}
\delta S_{t}=\frac{1}{8\pi G}\int{\sqrt{g}[R^{\lambda}_{\,\,\mu}-\frac{1}{2}\delta^{\lambda}_{\,\,\mu}R]e^{\alpha}_{\,\,\lambda}\delta e_{\alpha}^{\,\,\mu}\,d^{4}x}+\nonumber\\\lambda(\int{\sqrt{g}R(x)\,d^{4}x}){\frac{\partial F(Q)}{\partial Q}\int{\bar{\psi}\gamma^{\beta}\psi[e_{\beta}^{\,\,0}e^{\alpha}_{\,\,\mu}\delta e^{\,\,\mu}_{\alpha}+\delta e^{\,\,0}_{\beta}]\sqrt{g}\,d^{3}y}}\nonumber\\+\int{d^{4}x \sqrt{g}\,e^{\alpha}_{\,\,\mu}\delta e^{\,\,\mu}_{\alpha}\bar{\psi}(\frac{i}{2}\gamma^{\beta}\stackrel{\leftrightarrow}{D}_{\beta}-m)\psi}+\int{d^{4}x \sqrt{g}\,\delta e^{\,\,\mu}_{\beta}\bar{\psi}(\frac{i}{2}\gamma^{\beta}\stackrel{\leftrightarrow}{D}_{\mu})\psi}\nonumber\\+\lambda\frac{\partial F(Q)}{\partial Q}(\int{\sqrt{g}\delta \bar{\psi} e^{\,\,0}_{\alpha}\gamma^{\alpha}\psi\,d^{3}y})(\int{\sqrt{g}R(x)\,d^{4}x})+\int{d^{4}x\sqrt{g}}\delta\bar{\psi}(i\gamma^{\alpha}D_{\alpha}-m)\psi=0
\end{eqnarray}

So we have two equation, the first is:
\begin{equation}\label{eq:motion gravity Dirac 2}
(i\gamma^{\alpha}D_{\alpha}-m+\lambda \frac{\partial F(Q)}{\partial Q} e^{\,\,0}_{\alpha}\gamma^{\alpha}\delta(z^{0}-t_{0})(\int{d^{4}x\,R\sqrt{g}}))\psi=0
\end{equation}
which mean that again we have phase transformation $ \psi=e^{i\frac{\partial F(Q)}{\partial Q}\lambda(\int{\sqrt{g}R(x)\,d^{4}x})\theta (z^{0}-t_{0})}\psi_{D} $ that eliminates the $ \delta(z^{0}-t_{0}) $ term.\\
The second equation (after a modification of equation (\ref{eq:modification}) ) is:

\begin{eqnarray} \label{eq:ver tetrad Action 2}
-[R^{\lambda}_{\,\,\mu}-\frac{1}{2}\delta^{\lambda}_{\,\,\mu}R]e^{\alpha}_{\,\,\lambda}=8\pi G\lambda(\int{\sqrt{g}R(x)\,d^{4}x}) \bar{\psi}\gamma^{\beta}\psi\delta(y^{0}-t_{0})\frac{\partial F(Q)}{\partial Q}[e^{\,\,0}_{\beta} e^{\alpha}_{\,\,\mu}+\delta_{\beta}^{\alpha}\delta_{\mu}^{0}]\nonumber\\+8\pi G e^{\alpha}_{\,\,\mu} \bar{\psi}(\frac{i}{2} \gamma^{\alpha}\stackrel{\leftrightarrow}{D}_{\alpha}-m)\psi + 8\pi G \frac{i}{4}\bar{\psi}(\gamma^{\alpha}\stackrel{\leftrightarrow}{D}_{\mu}+\gamma_{\mu}\stackrel{\leftrightarrow}{D^{\alpha}})\psi
\end{eqnarray}

So, if we follow after the paragraph before and use the constraint of equation (\ref{eq: Strike on coordinate}) we will get:
\begin{eqnarray}\label{eq:Einstein field equation 4}
R_{\gamma\mu}-\frac{1}{2}g_{\mu\gamma}R=-8\pi G\bar{\psi}(\frac{i}{4} \gamma_{\gamma}\stackrel{\leftrightarrow}{D}_{\mu}+\frac{i}{4} \gamma_{\mu}\stackrel{\leftrightarrow}{D}_{\gamma})\psi\nonumber\\-8\pi \lambda G(\int{\sqrt{g}R(x)\,d^{4}x}) \bar{\psi}\gamma^{\beta}e^{\,\,\delta}_{\beta}g_{\delta 0}\psi\delta(y^{0}-t_{0})\frac{\partial F(Q)}{\partial Q}[\delta^{0}_{\gamma}\delta^{0}_{\mu}]
\end{eqnarray}

if we express equation (\ref{eq:Einstein field equation 4}) in term of $ \psi_{D} $ , then equation (\ref{eq: Strike on coordinate}) once again is obtained and finally , once again, we will get the Einstein equation (\ref{eq:Einstein field equation 3})

\section{Mass coupling constant as a general function in curved space}
The action of Dirac is:
\begin{equation}
S_{D}=\int{d^{4}x \sqrt{g}\bar{\psi}(\frac{i}{2}\gamma^{\alpha}\stackrel{\leftrightarrow}{D}_{\alpha}-m)\psi}
\end{equation} 

where we will defined the mass to be:
\begin{equation}\label{eq:def m curved}
m=\lambda\,F(\int{\sqrt{g}\bar{\psi}e^{\,\,0}_{\beta} \gamma^{\beta}\psi \,d^{3}y})=\lambda\,F(Q)
\end{equation}

and the action of the curved space is defined by equation (\ref{eq: coordinate action}),
so the varation of the total action is:

\begin{eqnarray} \label{eq: veration of gravity with mass}
\delta S_{t}=\frac{1}{8\pi G}\int{\sqrt{g}[R^{\lambda}_{\,\,\mu}-\frac{1}{2}\delta^{\lambda}_{\,\,\mu}R]e^{\alpha}_{\,\,\lambda}\delta e_{\alpha}^{\,\,\mu}\,d^{4}x}-\nonumber\\\lambda(\int{\sqrt{g}\bar{\psi}\psi\,d^{4}x}){\frac{\partial F(Q)}{\partial Q}\int{\bar{\psi}\gamma^{\beta}\psi[e_{\beta}^{\,\,0}e^{\alpha}_{\,\,\mu}\delta e^{\,\,\mu}_{\alpha}+\delta e^{\,\,0}_{\beta}]\sqrt{g}\,d^{3}y}}\nonumber\\+\int{d^{4}x \sqrt{g}\,e^{\alpha}_{\,\,\mu}\delta e^{\,\,\mu}_{\alpha}\bar{\psi}(\frac{i}{2}\gamma^{\beta}\stackrel{\leftrightarrow}{D}_{\beta}-m)\psi}+\int{d^{4}x \sqrt{g}\,\delta e^{\,\,\mu}_{\beta}\bar{\psi}(\frac{i}{2}\gamma^{\beta}\stackrel{\leftrightarrow}{D}_{\mu})\psi}\nonumber\\-\lambda\frac{\partial F(Q)}{\partial Q}(\int{\sqrt{g}\delta \bar{\psi} e^{\,\,0}_{\alpha}\gamma^{\alpha}\psi\,d^{3}y})(\int{\sqrt{g}\bar{\psi}\psi\,d^{4}x})+\int{d^{4}x\sqrt{g}}\delta\bar{\psi}(i\gamma^{\alpha}D_{\alpha}-m)\psi=0
\end{eqnarray}
as we done at the last section, we can have two equation:

\begin{equation}\label{eq:motion gravity Dirac with mass}
(i\gamma^{\alpha}D_{\alpha}-m-\lambda \frac{\partial F(Q)}{\partial Q} e^{\,\,0}_{\alpha}\gamma^{\alpha}\delta(z^{0}-t_{0})(\int{d^{4}x\,\bar{\psi}\psi\sqrt{g}}))\psi=0
\end{equation}
which mean that again we have a phase transformation $ \psi=e^{-i\frac{\partial F(Q)}{\partial Q}\lambda(\int{\bar{\psi}\psi\,d^{4}x})\theta (z^{0}-t_{0})}\psi_{D} $ that eliminates the $ \delta(z^{0}-t_{0}) $ term\\
and
\begin{eqnarray}\label{eq:Einstein field equation 5}
R_{\gamma\mu}-\frac{1}{2}g_{\mu\gamma}R=-8\pi G\bar{\psi}(\frac{i}{4} \gamma_{\gamma}\stackrel{\leftrightarrow}{D}_{\mu}+\frac{i}{4} \gamma_{\mu}\stackrel{\leftrightarrow}{D}_{\gamma})\psi\nonumber\\+8\pi \lambda G(\int{\sqrt{g}\bar{\psi}\psi\,d^{4}x}) \bar{\psi}\gamma^{\beta}\psi\delta(y^{0}-t_{0})\frac{\partial F(Q)}{\partial Q}[e^{\,\,\delta}_{\beta} g_{\delta\gamma}\delta^{0}_{\mu}]
\end{eqnarray}

we can see that again we need to use the condition on the coordinate as in equation (\ref{eq: Strike on coordinate})
so finally with the condition we will have:

\begin{eqnarray}\label{eq:Einstein field equation with mass}
R_{\gamma\mu}-\frac{1}{2}g_{\mu\gamma}R=-8\pi G\bar{\psi}(\frac{i}{4} \gamma_{\gamma}\stackrel{\leftrightarrow}{D}_{\mu}+\frac{i}{4} \gamma_{\mu}\stackrel{\leftrightarrow}{D}_{\gamma})\psi\nonumber\\+8\pi \lambda G(\int{\sqrt{g}\bar{\psi}\psi\,d^{4}x}) \bar{\psi}\gamma^{\beta}e^{\,\,\delta}_{\beta}g_{\delta 0}\psi\delta(y^{0}-t_{0})\frac{\partial F(Q)}{\partial Q}[\delta^{0}_{\gamma}\delta^{0}_{\mu}]
\end{eqnarray}

if we express equation (\ref{eq:Einstein field equation with mass}) in term of $ \psi_{D} $ , then equation (\ref{eq:motion gravity Dirac with mass}) once again is obtained and finally , once again, we will get the Einstein equation (\ref{eq:Einstein field equation 3})

\section{Cosmological constant depending on the total charge in the universe}
We will show that we can make a cosmological constant that depends on the total charge of the universe.
we will start with the action:

\begin{equation}
S_{D}=\int{d^{4}x \sqrt{g}\bar{\psi}(\frac{i}{2}\gamma^{\alpha}\stackrel{\leftrightarrow}{D}_{\alpha}-m)\psi}
\end{equation}  

where $ D_{\alpha}=e^{\,\,\mu}_{\alpha}\partial_{\mu}-\frac{i}{4}e^{\,\,\mu}_{\alpha}\eta_{\alpha c} \omega^{c}_{ b\mu}\sigma^{\alpha b} $ is the covariant derivative for fermion and $ e^{\mu}_{\, \alpha} $ is the vierbein and $ \sigma^{\alpha \beta}=\frac{i}{2}[\gamma^{\alpha},\gamma^{\beta}] $ is the commutator of the Dirac gamma metric, and the spin connection is $ \omega^{c}_{b\mu}=e^{c}_{\nu}\partial_{\mu}e^{\nu}_{b}+e^{c}_{\nu}e^{\sigma}_{b}\Gamma^{\nu}_{\sigma\mu} $ where $ \Gamma^{\nu}_{\sigma\mu} $ is the Christoffel symbol and $ \eta_{\alpha c} $ is the lorentzian metric.
the second definition is  the action of the gravity \cite{grav}

\begin{equation} \label{eq: coordinate action 2}
S_{G}=-\frac{1}{16\pi G}\int{\sqrt{g}(R(x)-\Lambda Q)\,d^{4}x}
\end{equation}
where $Q=\int\sqrt{g}\bar{\psi}e^{\,\,0}_{\alpha} \gamma^{\alpha}\psi \,d^{3}y $ and $ \int{\sqrt{g}\Lambda Q \,d^{4}x} $ is a charge dependent cosmological constant like term.
The total action is $ S_{t}=S_{D}+S_{G} $.
So by variation we will get:

\begin{eqnarray} \label{eq: veration of cosmological constant}
\delta S_{t}=\frac{1}{8\pi G}\int{\sqrt{g}[R^{\lambda}_{\,\,\mu}-\frac{1}{2}\delta^{\lambda}_{\,\,\mu}R]e^{\alpha}_{\,\,\lambda}\delta e_{\alpha}^{\,\,\mu}\,d^{4}x}+\nonumber\\\frac{\Lambda}{16\pi G}(\int{\sqrt{g}\,d^{4}x}){\int{\bar{\psi}\gamma^{\beta}\psi[e_{\beta}^{\,\,0}e^{\alpha}_{\,\,\mu}\delta e^{\,\,\mu}_{\alpha}+\delta e^{\,\,0}_{\beta}]\sqrt{g}\,d^{3}y}}+\frac{\Lambda Q}{16\pi G}\int{\sqrt{g}\,e^{\alpha}_{\,\,\mu}\delta e^{\,\,\mu}_{\alpha}\,d^{4}x}+\nonumber\\\int{d^{4}x \sqrt{g}\,e^{\alpha}_{\,\,\mu}\delta e^{\,\,\mu}_{\alpha}\bar{\psi}(\frac{i}{2}\gamma^{\beta}\stackrel{\leftrightarrow}{D}_{\beta}-m)\psi}+\int{d^{4}x \sqrt{g}\,\delta e^{\,\,\mu}_{\beta}\bar{\psi}(\frac{i}{2}\gamma^{\beta}\stackrel{\leftrightarrow}{D}_{\mu})\psi}\nonumber\\+\frac{\Lambda}{16\pi G}(\int{\sqrt{g}\delta \bar{\psi} e^{\,\,0}_{\alpha}\gamma^{\alpha}\psi\,d^{3}y})(\int{\sqrt{g}\,d^{4}x})+\int{d^{4}x\sqrt{g}}\delta\bar{\psi}(i\gamma^{\alpha}D_{\alpha}-m)\psi=0
\end{eqnarray}

because we have two different spaces, we can produce two equation. the first equation is:

\begin{equation} 
\frac{\Lambda}{16\pi G}(\int{\sqrt{g}\delta \bar{\psi} e^{\,\,0}_{\alpha}\gamma^{\alpha}\psi\,d^{3}y})(\int{\sqrt{g}\,d^{4}x})+\int{d^{4}x\sqrt{g}}\delta\bar{\psi}(i\gamma^{\alpha}D_{\alpha}-m)\psi=0
\end{equation}

we will use equation (\ref{eq:trick}) and the fact that $ \frac{\delta\bar{\psi}_{a}(x)}{\delta\bar{\psi}_{b}(z)}=\delta^{4}(x-z)\, \delta_{ab} $ and $ \frac{\delta\psi(x)}{\delta\bar{\psi}(z)}=0 $ and after integration we get:

\begin{equation}\label{eq:motion gravity Dirac cosmological}
(i\gamma^{\alpha}D_{\alpha}-m+\frac{\Lambda}{16\pi G} e^{\,\,0}_{\alpha}\gamma^{\alpha}\delta(z^{0}-t_{0})(\int{d^{4}x\,\sqrt{g}}))\psi=0
\end{equation}

which mean that again we have a gauge invariance transform which his solution is $ \psi=e^{i\frac{\Lambda}{16\pi G}(\int{\sqrt{g}\,d^{4}x})\theta (z^{0}-t_{0})}\psi_{D} $.

The second equation that follows by the variation is:
\begin{eqnarray} \label{eq:ver tetrad Action cosmological}
-[R^{\lambda}_{\,\,\mu}-\frac{1}{2}\delta^{\lambda}_{\,\,\mu}R+\frac{\Lambda Q}{2}\delta^{\lambda}_{\,\,\mu}]e^{\alpha}_{\,\,\lambda}=\frac{\Lambda}{2}(\int{\sqrt{g}\,d^{4}x}) \bar{\psi}\gamma^{\beta}\psi\delta(y^{0}-t_{0})[e^{\,\,0}_{\beta} e^{\alpha}_{\,\,\mu}+\delta_{\beta}^{\alpha}\delta_{\mu}^{0}]\nonumber\\+8\pi G e^{\alpha}_{\,\,\mu} \bar{\psi}(\frac{i}{2} \gamma^{\alpha}\stackrel{\leftrightarrow}{D}_{\alpha}-m)\psi + 8\pi G \frac{i}{4}\bar{\psi}(\gamma^{\alpha}\stackrel{\leftrightarrow}{D}_{\mu}+\gamma_{\mu}\stackrel{\leftrightarrow}{D^{\alpha}})\psi
\end{eqnarray}

By contracting equation (\ref{eq:ver tetrad Action cosmological}) with $ e_{\alpha\gamma}=e^{\delta}_{\,\,\gamma}\eta_{\alpha\delta} $ and use the fact that $ e^{\alpha}_{\,\,\lambda}e^{\delta}_{\,\,\gamma}\eta_{\alpha\delta}=g_{\lambda\gamma} $ and the fact that $ \bar{\psi}(\frac{i}{2} \gamma^{\alpha}\stackrel{\leftrightarrow}{D}_{\alpha}-m)\psi=-\frac{\Lambda}{16\pi G}(\int{\sqrt{g}\,d^{4}x}) \bar{\psi}e^{\,\,0}_{\beta}\gamma^{\beta}\psi\delta(y^{0}-t_{0}) $ we get the Einstein field equation with a modification 

\begin{eqnarray}\label{eq:Einstein field equation cosmological}
R_{\gamma\mu}-\frac{1}{2}g_{\mu\gamma}(R-\Lambda Q)=-8\pi G\bar{\psi}(\frac{i}{4} \gamma_{\gamma}\stackrel{\leftrightarrow}{D}_{\mu}+\frac{i}{4} \gamma_{\mu}\stackrel{\leftrightarrow}{D}_{\gamma})\psi\nonumber\\-\frac{\Lambda}{2}(\int{\sqrt{g}\,d^{4}x}) \bar{\psi}\gamma^{\beta}\psi\delta(y^{0}-t_{0})[e^{\,\,\delta}_{\beta} g_{\delta\gamma}\delta^{0}_{\mu}]
\end{eqnarray}

From here we follows the step as was done from equation (\ref{eq:Einstein field equation}) to equation (\ref{eq:Einstein field equation 3}) where now $ \psi=e^{i\frac{\Lambda}{16\pi G}(\int{\sqrt{g}\,d^{4}x})\theta (z^{0}-t_{0})}\psi_{D} $ and we have the Einstein equation with cosmological constant:

\begin{eqnarray}\label{eq:Einstein field equation cosmological 3}
R_{\gamma\mu}-\frac{1}{2}g_{\mu\gamma}(R-\Lambda Q)=-8\pi G\bar{\psi}_{D}(\frac{i}{4} \gamma_{\gamma}\stackrel{\leftrightarrow}{D}_{\mu}+\frac{i}{4} \gamma_{\mu}\stackrel{\leftrightarrow}{D}_{\gamma})\psi_{D}
\end{eqnarray}

we got a cosmological constant which depends on the charge of all the universe.

\section{Using approximately conserved charges, the large number hypothesis}
So far we have discussed the construction of the coupling constants as functions of conserved quantities, like electric charge, but there are many known charges that although our standard theories predict to be non conserved, there is still no direct evidence for it, for example baryon number (the matter anti matter asymmetry may indicate to a violation of baryon number \cite{baryon} in the early universe, but does not prove it) . If we were to use such non conserved charge, then we will not be able to restore Lorentz invariance, the time slice where the charge is defined will be important, the theory constructed in this way could behave very different than the standard one say in the early universe, but in more "calm" periods, where there is no even a direct hint of baryon number violation, the differences with the standard model would dissapear. \\

It is interesting also to note that  A. Eddington thought of the possibility that Newton's G could be influence by the large number of particles \cite{Eddington}. In our case indeed we can formulate a theory where G=G(N), where N is for example number of baryons, the theory will respect Lorentz invariance if we work under the approximation that baryon number is conserved, not exact in the standard model, but very close to exact indeed. 

\section{Coupling constant depending on Pontryagin index}
Until now we have defined the coupling constant to be a function of:
\begin{equation}
Q=\int{\bar{\psi}\gamma^{0}\psi\,d^{4}x}
\end{equation}
Now we will mention the possibility to anther direction to another direction which the coupling constants are function of the Pontryagin index or the "winding number" of some gauge fields:\cite{jackiw}
\begin{equation}
N=\frac{1}{16\pi^{2}} \int{d^{4}x\,Tr(\epsilon^{\mu\nu\alpha\beta}F_{\mu\nu}F_{\alpha\beta})}=\frac{1}{16\pi^{2}}\int{d^{4}x\,Tr(\partial_{\mu}(\epsilon^{\mu\nu\alpha\beta}( A_{\nu}F_{\alpha\beta}-\frac{2}{3}A_{\alpha}A_{\beta}A_{\nu})))}
\end{equation}
this number measures the number of time the hypersphere at infinity is wrapped around $ G_{\mu}=\epsilon^{\mu\nu\alpha\beta}( A_{\nu}F_{\alpha\beta}-\frac{2}{3}A_{\alpha}A_{\beta}A_{\nu}) $.
\\one can see that any local variation of $ A_{\mu} $ which leaves the boundaries unaffected douse not change N,therefore:
\begin{equation}
\delta N=0
\end{equation}
so now by construction of the coupling constant to be function of $ N $ we will get the ordinary Dirac equation because of the $ \delta N=0 $ situation.
Notice however that the winding number is only well defined for configurations of finite action in Euclidean space, like for example instantons.

\section{Non local contribution motivated from "infra red counter terms"}
In ref.\cite{jackiw2} ,in order to cancel infra red divergence in 3  - dimension gauge theories , infra red counter terms of the form:
\begin{equation}\label{eq: counter term}
c(\int j^{\mu}\, d^{3}y)(\int j_{\mu}\, d^{3}x)
\end{equation}
are introduced. This is shown in ref.\cite{jackiw2} to be equivalent to the procedure developed in ref\cite{guend} to cancel infra red divergence by introducing zero energy momentum photons.
These infra red divergences are related to the super renormalizabilty of the theory in 3- dimensions \cite{jackiw3}
We generalize this idea and define the coupling constant to be proportional to this term (and latter to be an arbitrary function of (\ref{eq: counter term})), we latter connect this procedure with our previous treatment.
\\We again begin by considering the action for the Dirac equation
\begin{equation}
S=\int d^{4}x\, \bar{\psi}(\frac{i}{2}\gamma^{\mu}\stackrel{\leftrightarrow}{\partial}_{\mu}-eA_{\mu}\gamma^{\mu}-m)\psi
\end{equation}
but now we take the coupling constant $ e $ to be proportional to the appropriate generalization of  (\ref{eq: counter term}) in the context of a four dimensional theory.

\begin{equation}\label{eq: e for jackiw}
e=\frac{\lambda}{2}(\int j^{\mu}\, d^{4}y)(\int j_{\mu}\, d^{4}z)=\frac{\lambda}{2}(\int\bar{\psi}\gamma^{\mu}\psi\, d^{4}y)(\int\bar{\psi}\gamma_{\mu}\psi\, d^{4}z)
\end{equation}

so the action will be

\begin{equation}
S=\int d^{4}x\, \bar{\psi}(x)(\frac{i}{2}\gamma^{\mu}\stackrel{\leftrightarrow}{\partial}_{\mu}-m)\psi(x)-\frac{\lambda}{2}(\int d^{4}x\, \bar{\psi}(x)A_{\mu}\gamma^{\mu}\psi(x))(\int\bar{\psi}(y)\gamma^{\mu}\psi(y)\, d^{4}y)(\int\bar{\psi}(z)\gamma_{\mu}\psi(z)\, d^{4}z)
\end{equation}

if we consider the fact that $ \frac{\delta\bar{\psi}_{a}(x)}{\delta\bar{\psi}_{b}(z)}=\delta^{4}(x-z)\,\delta_{ab} $ and $ \frac{\delta\psi(x)}{\delta\bar{\psi}(z)}=0 $ we get the equation of motion

\begin{eqnarray}
\frac {\delta S}{\delta\bar{\psi}(z)}=(i\gamma^{\mu}\partial_{\mu}-m)\psi(z)-\frac{\lambda}{2}(\int\bar{\psi}(z)\gamma^{\mu}\psi(z)\, d^{4}z)(\int\bar{\psi}(z)\gamma_{\mu}\psi(z)\, d^{4}z)A_{\mu}\gamma^{\mu}\psi(z)&\nonumber\\-\lambda(\int{\bar{\psi}(x)A_{\nu}\gamma^{\nu}\psi(x)}\,d^{4}x)(\int\bar{\psi}(z)\gamma^{\mu}\psi(z)\, d^{4}z)\gamma_{\mu}\psi(z)
\end{eqnarray}

which can be simplified more by the use of new definition  $ b^{\mu}_{e}=\lambda(\int{\bar{\psi}(x)A_{\nu}\gamma^{\nu}\psi(x)}\,d^{4}x)(\int\bar{\psi}(z)\gamma^{\mu}\psi(z)\, d^{4}z) $ which is a constant, and by the definition in equation (\ref{eq: e for jackiw}) 

\begin{equation}\label{eq:motion e 2}
\frac {\delta S}{\delta\bar{\psi}(z)}=[i\gamma^{\mu}\partial_{\mu}-m-eA_{\mu}\gamma^{\mu}-b^{\mu}_{e}\gamma_{\mu}]\psi(z)=0
\end{equation}

The solution of this equation is
\begin{equation}
\psi=e^{-ib^{\mu}_{e}x_{\mu}}\psi_{D} 
\end{equation}

where $ \psi_{D} $ is the solution of the equation
  \begin{equation}
 [i\gamma^{\mu}\partial_{\mu}-m-eA_{\mu}\gamma^{\mu}]\psi_{D}=0 
  \end{equation}
So we have just gauge transformation.  
We can connect this kind of contractions to what we did before in the page by generalizing the way we introduce the charge in the action from
\begin{equation}
e=\lambda\int{d^{4}x\delta(t-t_{0})\bar{\psi}\gamma^{0}\psi}
\end{equation}
to
\begin{equation}
e=\int{d^{4}x\sum_{i}\lambda_{i}\delta(t-t_{i})\bar{\psi}\gamma^{0}\psi} \\ \sum_{i}\lambda_{i}=\lambda
\end{equation}
and can use also instead of $ \sum_{i}\lambda_{i}\delta(t-t_{i}) $ an arbitrary function of time $ f(t) $
\begin{equation}
e=\int{f(t)\bar{\psi}(t,\textbf{x})\gamma^{0}\psi(t,\textbf{x})\,d^{4}x},\\ \int{f(t)\,dt}=\lambda
\end{equation}
from the special case $ f(t)=const $, we have to deal with
\begin{equation}\label{eq: connection to jackiw}
\int{\bar{\psi}\gamma^{0}\psi\,d^{4}x}=n_{\mu}\int{\bar{\psi}\gamma^{\mu}\psi\,d^{4}x}
\end{equation}
where $ n_{\mu}=(1,0,0,0) $.
Notice that in principle $ n_{\mu} $ can be an arbitrary constant vector, then all the procedure goes thought, i.e. the non covariant terms are possible to eliminate by a gauge transformation. The desired construction is obtained when choosing for $ n_{\mu} $ also $ \int{\bar{\psi}\gamma_{\mu}\psi\,d^{4}x} $ leading them to
\begin{equation}
\frac{1}{2}(\int{\bar{\psi}\gamma^{\mu}\psi\,d^{4}x})(\int{\bar{\psi}\gamma_{\mu}\psi\,d^{4}x})
\end{equation}
the $ \frac{1}{2} $ factor is included now, because $ n_{\mu} $ has become dynamical and its variation also contribute to the equations of motion.
\section{Conclusion}
We have constructed a model that showed that we can make a new "Mach principle" for charge, in which all the electromagnetic coupling constant are a function of the total charge in the universe.

\begin{equation}
Q=\int\psi^{\dagger}(\vec{y},y^{0}=t_{0})\psi(\vec{y},y^{0}=t_{0}) \,d^{3}y=\int \rho(\vec{y},y^{0}=t_{0}) \,d^{3}y
\end{equation}

We have showed that Lorentz invariance can be restored by just a phase transformation of the Dirac field.
This shows that coupling constants can be taken not just as a given number but as a global function that depends on the global state of the universe in particular on the total charge of the universe, in the context of a consistent formalism.
This is an explicit realization of something similar to the Mach principle, but now applied to charge rather than to mass.
\\Although we have considered a charge that has been coupled to a gauge field, this is not necessary. We could consider also a charge which is not gauged, like for example baryon number, although baryon number in the standard model is not strictly conserved, then in a theory where the phase of baryons can be changed globally leaving the action invariant, we could also consider the coupling constants as a function of total baryon number. The apparent non Lorentz invariance can once again be cancelled by a phase transformation.
however the method is guaranteed to work only in the case where baryon number is assumed to be conserved, any violation of baryon number \cite{baryon} will induce a corresponding violation of Lorentz invariance 
Furthermore, when we work in curved spacetime we can see that we have gauge constraint on the coordinate, in which the coordinate prevent any local current:

\begin{equation}
\bar{\psi}\gamma^{\beta}e^{\,\,\delta}_{\beta}g_{\delta i}\psi=0
\end{equation}

where the reason is that we need that the Einstein equation will be symmetric.
Although after the coordinate constraint is applied we can change the coordinate as we wants.\\
We can also follow the method and make a cosmology constant that depends on the total charge of the universe.\\One of our objectives was to make a no local theory, so we have also shown that we can also make a theory that the coupling constant is a function of Pontryagin index.
In ref.\cite{jackiw2} ,in order to cancel infra red divergence in 3  - dimension gauge theories , infra red counter terms are introduced. This is shown in ref.\cite{jackiw2} to be equivalent to the procedure developed in ref\cite{guend} to cancel infra red divergence by introducing zero energy momentum photons.
These infra red divergences are related to the super renormalizabilty of the theory in 3- dimensions \cite{jackiw3}
We generalize this idea and define the coupling constant to be proportional to this term (and latter to be an arbitrary function of (\ref{eq: counter term})), we have connected this procedure with our treatment.

\appendix

\section{The Lorentz invariant of Q by the conservation of the 4 current}\label{appendix Q}
We will define the four Dirac current to be:
\begin{equation}
j^{\mu}=\bar{\psi}(y)\gamma^{\mu}\psi(y)
\end{equation}
and since we know after we use the equation of motion that the four dimension current is conserved:
\begin{equation}\label{conservation of current}
\partial_{\mu}j^{\mu}=0
\end{equation}
the total charge is:
\begin{equation}
Q=\int{j^{0}(x)\,d^{3}x}
\end{equation}
following Weinberg \cite{grav} we can rewrite the charge as:
\begin{equation}
Q=\int{d^{4}x\,j^{\mu}\partial_{\mu}\theta(n^{\nu}x_{\nu})}
\end{equation}
where $ \theta(a) $ is a step function, and by definition $ n_{i}=0\, , \, n_{0}=1 $.
If we perform a Lorentz transformation on $ Q $ then:
\begin{equation}
Q'=\int{d^{4}x'\,j'^{\mu}(x')\partial'_{\mu}\theta(n^{\nu}x'_{\nu})}=\int{d^{4}x\,\Lambda^{\sigma}_{\,\,\mu} j^{\mu}\Lambda^{\,\,\tau}_{\sigma}\partial_{\tau}\theta(n^{\nu}x'_{\nu})}=\int{d^{4}x\,j^{\mu}\partial_{\mu}\theta(n^{\nu}x'_{\nu})}
\end{equation}
where $ n_{\lambda}x'^{\lambda}=n'_{\rho}x^{\rho} $
so
\begin{equation}
Q'=\int{d^{4}x\,j^{\mu}\partial_{\mu}\theta(n'^{\nu}x_{\nu})}
\end{equation}
if we using equation (\ref{conservation of current}) then we can write:
\begin{equation}
Q'-Q=\int{d^{4}x\,\partial_{\mu}[j^{\mu}\lbrace \theta(n'_{\nu}x^{\nu})-\theta(n_{\nu}x^{\nu}) \rbrace]}
\end{equation}
The current $ j^{\mu} $ can be presumed to vanish if $ \vert \textbf{x} \vert \longrightarrow \infty $ with t fixed, whereas the function $  \theta(n'_{\nu}x^{\nu})-\theta(n_{\nu}x^{\nu}) $ vanishes if $ \vert t \vert \longrightarrow\infty $ with \textbf{x} fixed, so $ Q'-Q=0 $ so the charge is invariant under Lorentz transform because $ n_{\nu} $ and $ n'_{\nu} $ is time like.
\section{The Klein Gordon and Schrodinger cases }
\subsection*{The Klein Gordon case}

Until now we have steadied the Dirac field, now we will show how these affects look in the Klein Gordon case.

\subsection*{When the Mass is a function of all the charge in the universe on the Klein Gordon field}
We will begin with the action of Klein Gordon equation:
\begin{equation}
S=\int{d^{4}x\,[(i\partial_{\mu}\phi^{*} +eA_{\mu}\phi^{*})(-i\partial^{\mu}\phi +eA^{\mu}\phi)-m^{2}\phi^{*}\phi]}
\end{equation}
where we will take:
\begin{equation}\label{eq: def m KG}
m^{2}=\lambda\int{d^{3}y\,[\phi^{*}i\stackrel{\leftrightarrow}{\partial^{0}}\phi -2eA^{0}\phi^{*}\phi]}=\lambda\int{d^{4}y\,[\phi^{*}i\stackrel{\leftrightarrow}{\partial^{0}}\phi -2eA^{0}\phi^{*}\phi]\delta(y^{0}-t_{0})}
\end{equation}
which is the total charge in the universe by the definition of Klein Gordon field.
So by variation we will get:
\begin{eqnarray}
\delta S=\int{d^{4}x\,[-\delta\phi^{*}\,\partial_{\mu}\partial^{\mu}\phi - i\delta\phi^{*} e\partial_{\mu}(A^{\mu}\phi)-i \delta\phi^{*} eA_{\mu}\partial^{\mu}\phi+e^{2}\delta\phi^{*}A_{\mu}A^{\mu}\phi-\delta\phi^{*} m^{2}\phi]}\nonumber\\-\lambda(\int{d^{4}x\,\phi^{*}\phi})\int{d^{4}y\,\delta(y^{0}-t_{0})[\delta\phi^{*}i\partial_{0}\phi+\delta\phi^{*}i\partial_{0}\phi-2e\delta\phi^{*}A^{0}\phi]}\nonumber\\
-\lambda(\int{d^{4}x\,\phi^{*}\phi})\int{d^{4}y\,\delta\phi^{*}i\phi \,\partial_{0}\delta(y^{0}-t_{0})}
\end{eqnarray}
from this we get the equation of motion:

\begin{eqnarray}\label{eq:equation motion Klein Gordon}
-\partial_{\mu}\partial^{\mu}\phi-ie\phi\partial_{\mu}A^{\mu}-2ieA^{\mu}\partial_{\mu}\phi+e^{2}A_{\mu}A^{\mu}\phi-m^{2}\phi-2\lambda(\int{d^{4}x\,\phi^{*}\phi})\delta(y^{0}-t_{0})[i\partial_{0}\phi-eA_{0}\phi]\nonumber\\-i\lambda(\int{d^{4}x\,\phi^{*}\phi})\phi\,\partial_{0}\delta(y^{0}-t_{0})=0
\end{eqnarray}

if we do the transformation 

\begin{equation}\label{eq:Klein A transform}
A^{0}\longrightarrow A^{0}+\frac{i\lambda_{1}b}{e}\delta(y^{0}-t_{0})
\end{equation}
and
\begin{equation}\label{eq:Klein psi transform}
\phi=e^{\lambda_{2} b\theta(y^{0}-t_{0})}\phi_{0}
\end{equation}
where $ b=i\lambda\int{\phi^{*}\phi\,d^{4}y} $
then we have that
\begin{eqnarray}
\partial_{0}\phi=e^{\lambda_{2}b\theta(y^{0}-t_{0})}\partial_{0}\phi_{0}+\lambda_{2}b\delta(y^{0}-t_{0})e^{\lambda_{2}b\theta(y^{0}-t_{0})}\phi_{0}
\\ \nonumber\\
\partial_{0}\partial^{0}\phi=e^{\lambda_{2}b\theta(y^{0}-t_{0})}\partial_{0}\partial^{0}\phi_{0}+2\lambda_{2}b\delta(y^{0}-t_{0})e^{\lambda_{2}b\theta(y^{0}-t_{0})}\partial_{0}\phi_{0}\nonumber\\+\lambda_{2}b^{2}\delta^{2}(y^{0}-t_{0})e^{\lambda_{2}b\theta(y^{0}-t_{0})}\phi_{0}+\lambda_{2}b(\partial_{0}\delta(y^{0}-t_{0}))e^{b\theta(y^{0}-t_{0})}\phi_{0}
\end{eqnarray}
So, if we use equation (\ref{eq:Klein A transform}) in equation (\ref{eq:equation motion Klein Gordon}) than we will get:

\begin{eqnarray}\label{eq:equation motion Klein Gordon with A TRNS}
-\partial_{\mu}\partial^{\mu}\phi-ie\phi\partial_{\mu}A^{\mu}-2ieA^{\mu}\partial_{\mu}\phi+e^{2}A_{\mu}A^{\mu}\phi-m^{2}\phi+2b\delta(y^{0}-t_{0})[(\lambda_{1}-1)\partial_{0}\phi\nonumber\\+i(\lambda_{1}-1)eA_{0}\phi+0.5b\delta(y^{0}-t_{0})\phi(2\lambda_{1}-\lambda^{2}_{1})]\nonumber\\+b(\lambda_{1}-1)\phi\,\partial_{0}\delta(y^{0}-t_{0})=0
\end{eqnarray}

and finally we put equation (\ref{eq:Klein psi transform}) and we get:

\begin{eqnarray}\label{eq:equation motion Klein Gordon 2}
-\partial_{\mu}\partial^{\mu}\phi_{0}-ie\phi_{0}\partial_{\mu}A^{\mu}-2ieA^{\mu}\partial_{\mu}\phi_{0}+e^{2}A_{\mu}A^{\mu}\phi_{0}-m^{2}\phi_{0}+2b\delta(y^{0}-t_{0})[(\lambda_{1}+\lambda_{2}-1)\partial_{0}\phi_{0}\nonumber\\+i(\lambda_{1}+\lambda_{2}-1)eA_{0}\phi_{0}+0.5b\delta(y^{0}-t_{0})\phi_{0}(\lambda^{2}_{2}+(2\lambda_{1}-\lambda^{2}_{1})+2(\lambda_{1}-1)\lambda_{2})]\nonumber\\+b(\lambda_{1}+\lambda_{2}-1)\phi_{0}\,\partial_{0}\delta(y^{0}-t_{0})=0
\end{eqnarray}
if we need that equation (\ref{eq:equation motion Klein Gordon 2}) will be like ordinary Klein Gordon equation we need that:

\begin{eqnarray}
\lambda_{1}+\lambda_{2}-1=0\\
\lambda^{2}_{2}+(2\lambda_{1}-\lambda^{2}_{1})+2(\lambda_{1}-1)\lambda_{2}=0
\end{eqnarray}

for which the solutions are $ \lambda_{2}=\frac{1}{\sqrt{2}} \,,\, -\frac{1}{{\sqrt{2}}} \, and \, \lambda_{1}=1-\frac{1}{\sqrt{2}} \,,\, 1+\frac{1}{\sqrt{2}}$ respectively

where for these solutions equation (\ref{eq:equation motion Klein Gordon 2}) will become:

\begin{equation}
[(i\partial_{\mu}-eA_{\mu})(i\partial^{\mu}-eA^{\mu})-m^{2}]\phi_{0}=0
\end{equation}

\subsection*{Coupling constant depending of Charge in Klein Gordon}
We will begin with the action of Klein Gordon equation:
\begin{equation}
S=\int{d^{4}x\,[(i\partial_{\mu}\phi^{*} +eA_{\mu}\phi^{*})(-i\partial^{\mu}\phi +eA^{\mu}\phi)-m^{2}\phi^{*}\phi]}
\end{equation}
where we will take:
\begin{equation}
e=\lambda\int{d^{3}y\,[\phi^{*}i\stackrel{\leftrightarrow}{\partial^{0}}\phi -2eA^{0}\phi^{*}\phi]}=\lambda\int{d^{4}y\,[\phi^{*}i\stackrel{\leftrightarrow}{\partial^{0}}\phi -2eA^{0}\phi^{*}\phi]\delta(y^{0}-t_{0})}
\end{equation}
which is the total charge in the universe by the definition of Klein Gordon.
So by variation we will get:
\begin{eqnarray}
\delta S=\int{d^{4}x\,[-\delta\phi^{*}\,\partial_{\mu}\partial^{\mu}\phi - i\delta\phi^{*} e\partial_{\mu}(A^{\mu}\phi)-i \delta\phi^{*} eA_{\mu}\partial^{\mu}\phi+e^{2}\delta\phi^{*}A_{\mu}A^{\mu}\phi-\delta\phi^{*} m^{2}\phi]}\nonumber\\-\lambda(\int{d^{4}x\,A_{\mu}(\phi^{*}i\stackrel{\leftrightarrow}{\partial^{\mu}}\phi -2eA^{\mu}\phi^{*}\phi)})\int{d^{4}y\,\delta(y^{0}-t_{0})[\delta\phi^{*}i\partial_{0}\phi+\delta\phi^{*}i\partial_{0}\phi-2e\delta\phi^{*}A^{0}\phi]}\nonumber\\
-\lambda(\int{d^{4}x\,A_{\mu}(\phi^{*}i\stackrel{\leftrightarrow}{\partial^{\mu}}\phi -2eA^{\mu}\phi^{*}\phi)})\int{d^{4}y\,\delta\phi^{*}i\phi \,\partial_{0}\delta(y^{0}-t_{0})}
\end{eqnarray}

where we know that $ \int{d^{4}x\,A_{\mu}(\phi^{*}i\stackrel{\leftrightarrow}{\partial^{\mu}}\phi -2eA^{\mu}\phi^{*}\phi)}=\int{d^{4}x\,A_{\mu}j^{\mu}} $ so the equation of motion will become:

\begin{eqnarray}\label{eq:equation motion Klein Gordon charge}
-\partial_{\mu}\partial^{\mu}\phi-ie\phi\partial_{\mu}A^{\mu}-2ieA^{\mu}\partial_{\mu}\phi+e^{2}A_{\mu}A^{\mu}\phi-m^{2}\phi-2\lambda(\int{d^{4}x\,A_{\mu}j^{\mu}})\delta(y^{0}-t_{0})[i\partial_{0}\phi-eA_{0}\phi]\nonumber\\-i\lambda(\int{d^{4}x\,A_{\mu}j^{\mu}})\phi\,\partial_{0}\delta(y^{0}-t_{0})=0
\end{eqnarray}

the solution of the equation is the same as in the last section but with $ b=i\lambda\int{d^{4}x\,A_{\mu}j^{\mu}} $

\subsection*{The Schrodinger case}
We can see that also in Schrodinger equation case these effects can be studied.
The action for the Schrodinger equation is \cite{basic}:

\begin{equation}
S=\int(\frac{1}{2}\psi^{*} i \stackrel{\leftrightarrow}{\partial_{t}}\psi-\frac{1}{2m}(\nabla\psi^{*}-e\textbf{A})(\nabla\psi-e\textbf{A})+eV\psi^{*}\psi)\,d^{4}x
\end{equation}

we can take the coupling constant as "e" and "m" and follow the same approach as before where the coupling constant is proportional to the total charge in the universe , for example \cite{basic 2}:

\begin{equation}
e(Q)=\lambda_{1}\int{\psi^{*}\psi\,d^{3}x}
\end{equation}
and
\begin{equation}
m(Q)=\lambda_{2}\int{\psi^{*}\psi\,d^{3}x}
\end{equation}
where
\begin{equation}
Q=\int{\psi^{*}\psi\,d^{3}x}=\int{\rho\,d^{3}x}
\end{equation}

The equation of motion will be like the standard Schrodinger equation, if we will preform the transformation:
\begin{equation}
\psi=e^{i(b_{e}+b_{m})\theta(t-t^{0})}\psi_{0}
\end{equation}

where $ \psi_{0} $ is the solution of the standard Schrodinger equation, and "b" is
\begin{equation}
b_{e}=\lambda_{1}\int{ ((\nabla\psi^{*})\textbf{A}+\textbf{A}(\nabla\psi)+2e\textbf{A}^{2}+\psi^{*}\psi V)\,d^{4}x}
\end{equation}
and
\begin{equation}
b_{m}=\frac{1}{2m^{2}}\lambda_{2}\int{ (\nabla\psi^{*}-e\textbf{A})(\nabla\psi-e\textbf{A})\,d^{4}x}
\end{equation}


\begin{thebibliography} {}
\bibitem {Mach}
 E.Mach.
\emph{The science of mechanics; a critical and historical account of its development \\ ( Chicago : The Open Court Publishing Co , 1915).}

\bibitem {basic}
 G.Franz
\emph{Relativistic Quantum Mechanics and Field Theory (John Wiley and Sons ,1993).}

\bibitem {dirac rel}
G. G. Nyambuya
\emph{New Curved Spacetime Dirac Equations (Foundations of Physics Journal, July 2008, Vol. 38, Issue 7, pages 665-677)}
\bibitem {grav}
S.Weinberg
\emph{Gravitation and Cosmology (1972)}
T.Ortin
\emph{Gravity and strings (2004)} 

\bibitem {baryon}
 M.Y.Khlopov
\emph{Cosmoparticle Physics (World scientific ,1999).}
\bibitem {Eddington}
 A. Eddington
 \emph{Preliminary Note on the Masses of the Electron, the Proton, and the Universe (Proceedings of the Cambridge Philosophical Society 27, 1931).}
 J.D.Barrow and F.J.Tipler
 \emph{The Anthropic cosmological Principle(Oxford University Press, 1986)}
\bibitem {jackiw}
{Belavin}, A.~A. and {Polyakov}, A.~M. and {Schwartz}, A.~S. and 
	{Tyupkin}, Y.~S.
 \emph{Pseudoparticle solutions of the Yang-Mills equations(Physics Letters B, 1975).}
See for review S.Coleman
 \emph{Aspects of symmetry (Cambridge university press, 1985)}
 \bibitem {basic 2}
 J.J.Sakurai
\emph{Modern Quantum Mechanics (Addison-Wesley ,1993).}
 
\bibitem {jackiw2}
 R.Jackiw
\emph{Gauge theories in three dimension (MIT CPT 1000, Arctic school of physics Akaslompolo Finland, International symposium on gauge theory and gravitation, Nara Japan, 3rd Marcel Grossmann meeting on general relativity Shanghai China, Symposium on high Energy physics Tokyo Japan ,1982).} 

\bibitem {jackiw3}
 R.Jackiw and S.Templeton
\emph{How super-renormalizable interactions cure their infrared divergences (phys. rev ,1981).} G.t Hooft \emph{ Field Theory and Strong Interactions(proceedings of the XIX Internationale Universitatswochen fur Kernphysik, Schladming (Acta Phys. Austriaca. Suppl. 22), edited by P. Urban (Springer, Vienna, 1980)}

\bibitem {guend}
 E.I.Guendelman and Z.Radulovic
\emph{Loop expansion in $ QED_{3} $ ( Phys. Rev. D27, 357-365, 1983).} , E.I.Guendelman and Z.Radulovic \emph{Infrared divergences in three dimensional gauge theories ( Phys. Rev. D30, 1338-1349, 1984).}
\end{thebibliography}
\end{document}